# Fock space formulation for nanoscale electronic transport

**Supriyo Datta**
*School of Electrical & Computer Engineering,*
*Purdue University,*
*West Lafayette, IN 47907, USA.*

**(March 1, 2006)**

## Abstract

In this paper we present a general formulation for electronic transport that combines strong correlation effects with broadening and quantum coherence, and illustrate it with a simple example ("spin blockade") that clearly demonstrates all three effects. The standard master equation method for Coulomb blockade captures the first effect, while a number of different approaches such as the non-equilibrium Green function (NEGF) method are available for handling the last two. But there is a need for a method that captures all three and in this paper we propose an approach that appears to fulfill this need. The equations look similar to the NEGF formalism, but one basic distinction is that all quantities (like the Hamiltonian) appearing in our formalism are defined in Fock space and as such are matrices of dimension ($2^N$ x $2^N$), N being the number of basis functions describing the one-electron Hilbert space. Similar quantities appearing in the NEGF formalism are of dimension (NxN). Other important differences arise from the fact that in Fock space there is no need to include the exclusion principle explicitly and that elastic processes in one-electron space look like inelastic processes in Fock space. A simple numerical example is presented to show that our approach includes strong electron-electron interaction, broadening due to coupling to contacts and quantum coherence effects due to spin polarization in directions other than the z-direction. The basic equations for this "Fock space Green's function" (FSGF) approach are quite general and should be applicable to more interesting and exotic transport problems involving the entanglement of many-electron states. How effective it will be remains to be assessed for different problems, but our preliminary results for the Kondo peak look encouraging.

## 1. Introduction

Quantum transport in nanostructures is a topic of great current interest and much progress has been made by combining the non-equilibrium Green function (NEGF) method with the physical insights from the Landauer approach, using an appropriate self-consistent field (SCF) to account for electron-electron interactions. However, these approaches treat interactions as a perturbation, making it difficult to include the effect of strong correlations, although possible refinements have been proposed [1]. Such strong correlations are known to lead to observable effects in the current-voltage (I-V) characteristics that cannot be described by standard SCF treatments [2].

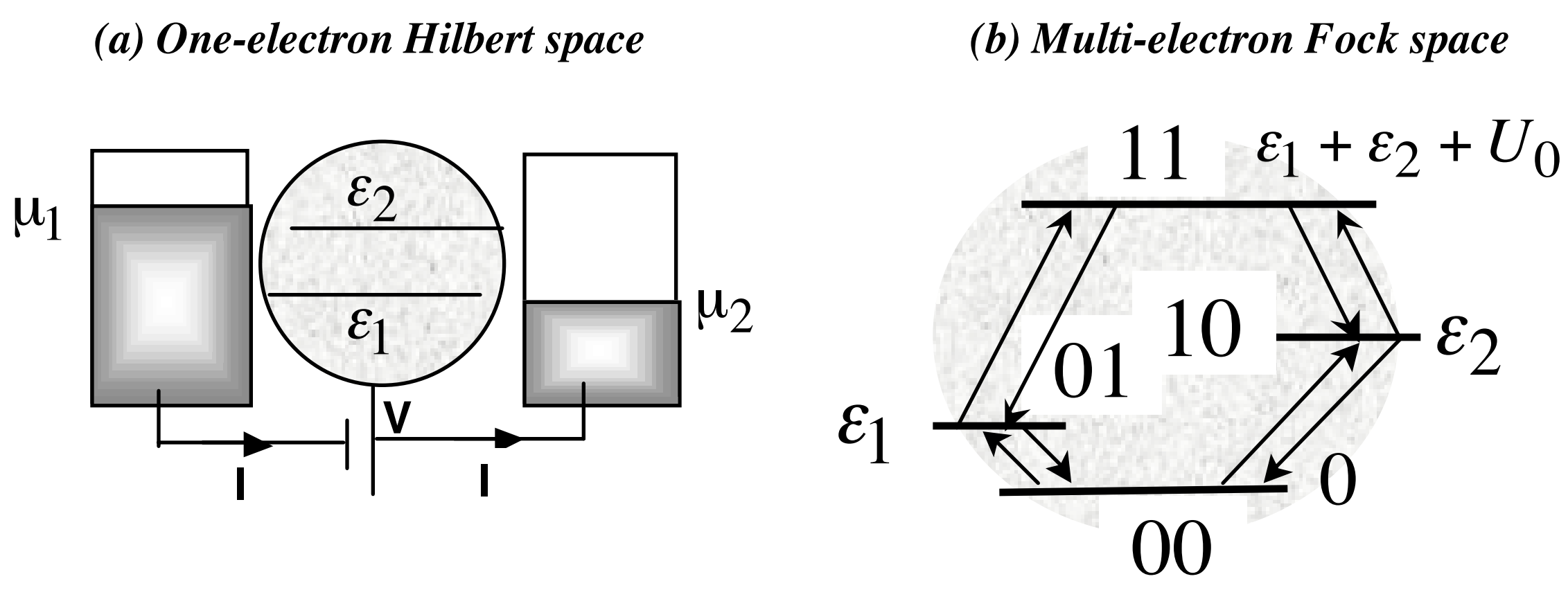

***Fig.1.1. (a) Transport problems are commonly formulated in the one-electron Hilbert space with N basis functions; N=2 is used in the illustration. (b) In the Coulomb blockade (CB) regime transport problems are described by a master equation using the $2^N$ energy eigenstates in Fock space as the basis functions. This paper presents a general formulation in the full $2^N$ dimensional Fock space that accounts for off-diagonal terms as well as broadening and as such does not require the use of basis functions that are eigenstates.***

Consider for example a small quantum dot having just two energy levels (see Fig.1.1a) in the energy range of interest, so that the relevant one-electron Hilbert space is of size two (N=2) and in an NEGF formulation the device would be described by (2x2) matrices. It is well-known that the usual SCF-NEGF approach is not accurate if the electron-electron interaction energy (per electron) $U_0$ is large compared to other energy scales such as the level broadening due to contacts / scattering / temperature. In this Coulomb blockade (CB) regime, the usual approach is to solve a master equation or a rate equation for the occupation probabilities of different states in the multi-electron Fock space (same as the configuration space in quantum chemistry), which is of size $2^N$ (= 4, if N=2) as shown in Fig.1.1b. Electron exchange processes with the reservoirs or contacts cause transitions among different states in Fock space whose electron number differ by one and the steady-state occupation probabilities are obtained by equating the transition rates into and out of each state [3]. This approach has been widely used to model the Coulomb blockade (CB) regime [4], and has been extended to include "co-tunneling"[5]. But it seems limited by the fact that it is based on the occupation probabilities of discrete states and thus cannot account for quantum interference and broadening of the type that are easily included in the NEGF approach. There has been some work on including quantum interference [6] through degenerate states but without accounting for the broadening that accompanies the coupling to the contacts. Several recent papers have described attempts to include energy level broadening into the Fock space description [7, 8] and could provide viable alternative approaches to the problem we are addressing in this paper. But at this time it is difficult to assess how these methods compare with the one presented here, which is very different from all other approaches that we are aware of.

The purpose of this paper is to present a general Fock space transport formalism combining strong correlation effects with quantum interference and broadening and to illustrate it with a simple example ("spin blockade") that clearly demonstrates quantum interference through energy levels that are not exactly degenerate to start with, but overlap in energy due to broadening. For example, Fig.1.2 shows the current-voltage (I-V) characteristics we calculate using the present formulation for a device with two spin-degenerate levels, with different values of the coupling of the device to the contacts described by the corresponding level broadening parameters $\gamma_1$ and $\gamma_2$. Note how the characteristics evolve from the staircase structure ($\gamma_1 = \gamma_2 = 0.2$ meV) typical of the CB regime to the smoother curves characteristic of the SCF regime ($\gamma_1 = \gamma_2 = 5$ meV). While this smooth transition is reassuring, it does not provide any indication that quantum coherence is being correctly described since the off-diagonal elements of the density matrix are essentially zero and play no role in the calculations.

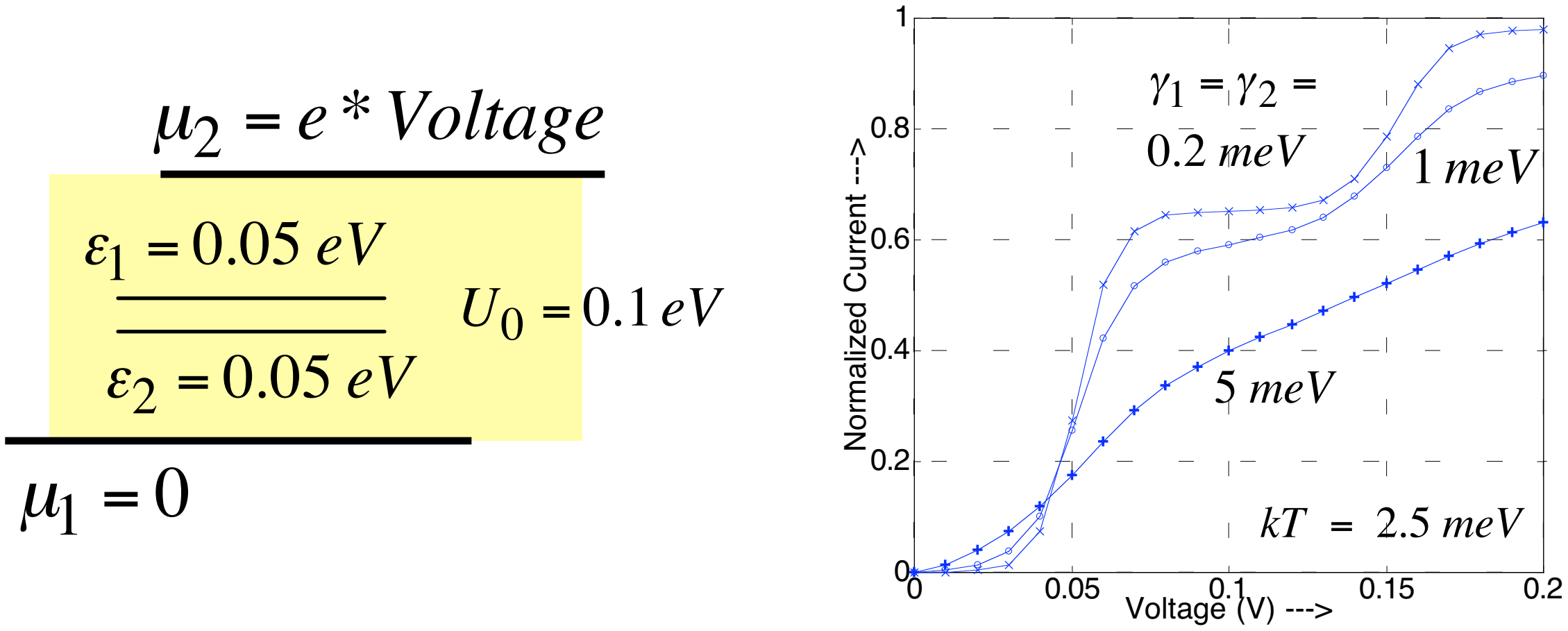

***Fig.1.2. Current-voltage (I-V) characteristics calculated using the present formulation for a device with two spin-degenerate level with different values of the coupling of the device to the contacts described by the corresponding level broadening parameters*** $\gamma_1$ ***and*** $\gamma_2$***. Current is normalized to*** $(e/\hbar)2\gamma_1\gamma_2/(\gamma_1+\gamma_2)$.

To illustrate this aspect we choose the more general problem of a quantum dot with two spin levels (not necessarily degenerate) having four contacts as shown in Fig.1.3. If $\gamma_3 = \gamma_1$ and $\gamma_4 = \gamma_2$ then both contacts are effectively unpolarized and we obtain characteristics like those shown in Fig.1.2. But if at least one of the contacts (say the left contact) is polarized so that $\gamma_3 \ll \gamma_1$ then the I-V characteristics can display a rich set of possibilities including a simple one quantum dot version of the ***spin blockade*** phenomenon recently observed in double quantum dots [9]. To describe such phenomena correctly, one requires a model that includes both strong electron-electron interaction and energy level broadening. Moreover, if the polarization angle $\theta$ is different from zero, then off-diagonal elements of the density matrix play an important role in determining the I-V characteristics thus providing a stringent check for processes involving quantum coherence.

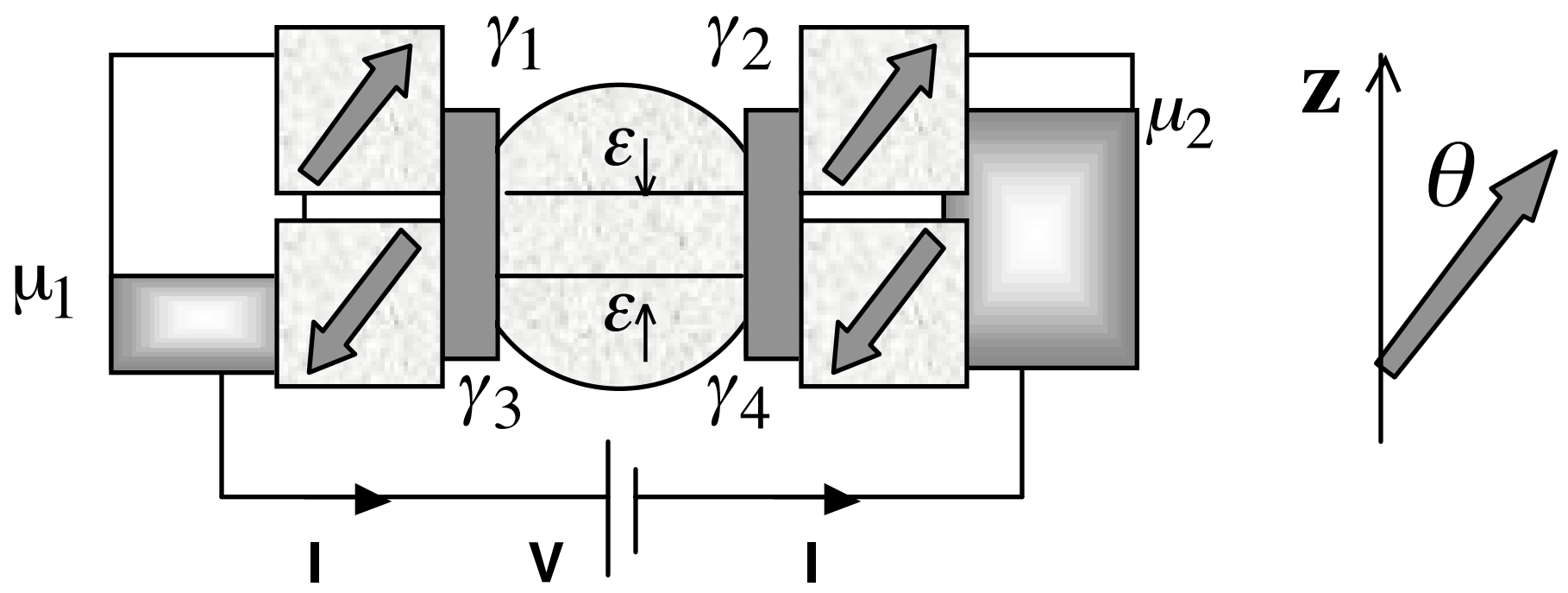

***Fig.1.3. I-V characteristics for this structure with four distinct spin-polarized contacts are used in Section 3 to illustrate our formalism including the role of strong electron-electron interactions, level broadening and quantum interference.***

In Section 3, we will illustrate our formulation by using it to analyze the structure shown in Fig.1.3 which we have chosen because it provides a simple embodiment of the essential ingredients we are hoping to capture, namely, strong interaction, broadening and quantum coherence. But although our choice is pedagogically motivated, we do predict interesting interference effects in the strongly correlated transport regime that could be checked experimentally using existing quantum dot devices. We believe that our approach could also be useful in describing recent experiments in double quantum dots with singlet-triplet correlations, as well as in investigating more exotic entangled states, but these applications are beyond the scope of the present paper. Our purpose here is primarily to present a general compact formulation of the approach (see ***Section 2***) along with a detailed derivation (see ***Appendix***). MATLAB codes for the numerical examples presented in ***Section 3*** are available on request.

## 2. Method

### 2.1. Model

Fig.2.1 illustrates the generic problem we wish to address. A strongly interacting "system" described by (indices include spin)

$$H \;=\; \sum_{i,j} h_{ij}\, c_i^{+} c_j \;+\; \sum_{i,j,k,l} U_{ijkl}\, c_i^{+} c_j^{+} c_l\, c_k \qquad (2.1)$$

and two (or more) "contacts" or reservoirs described by

$$H_1 \;=\; \sum_{r,s\in 1} h_{rs}\, c_r^{+} c_s \quad , \quad H_2 \;=\; \sum_{r,s\in 2} h_{rs}\, c_r^{+} c_s \qquad (2.2)$$

coupled through electron exchange processes:

$$\tau_1 = \sum_{r\in 1\,,\,j} t_{rj}\, c_r^{+} c_j \;+\; t_{jr}\, c_j^{+} c_r$$

$$\tau_2 = \sum_{r\in 2\,,\,j} t_{rj}\, c_r^{+} c_j \;+\; t_{jr}\, c_j^{+} c_r \qquad (2.3)$$

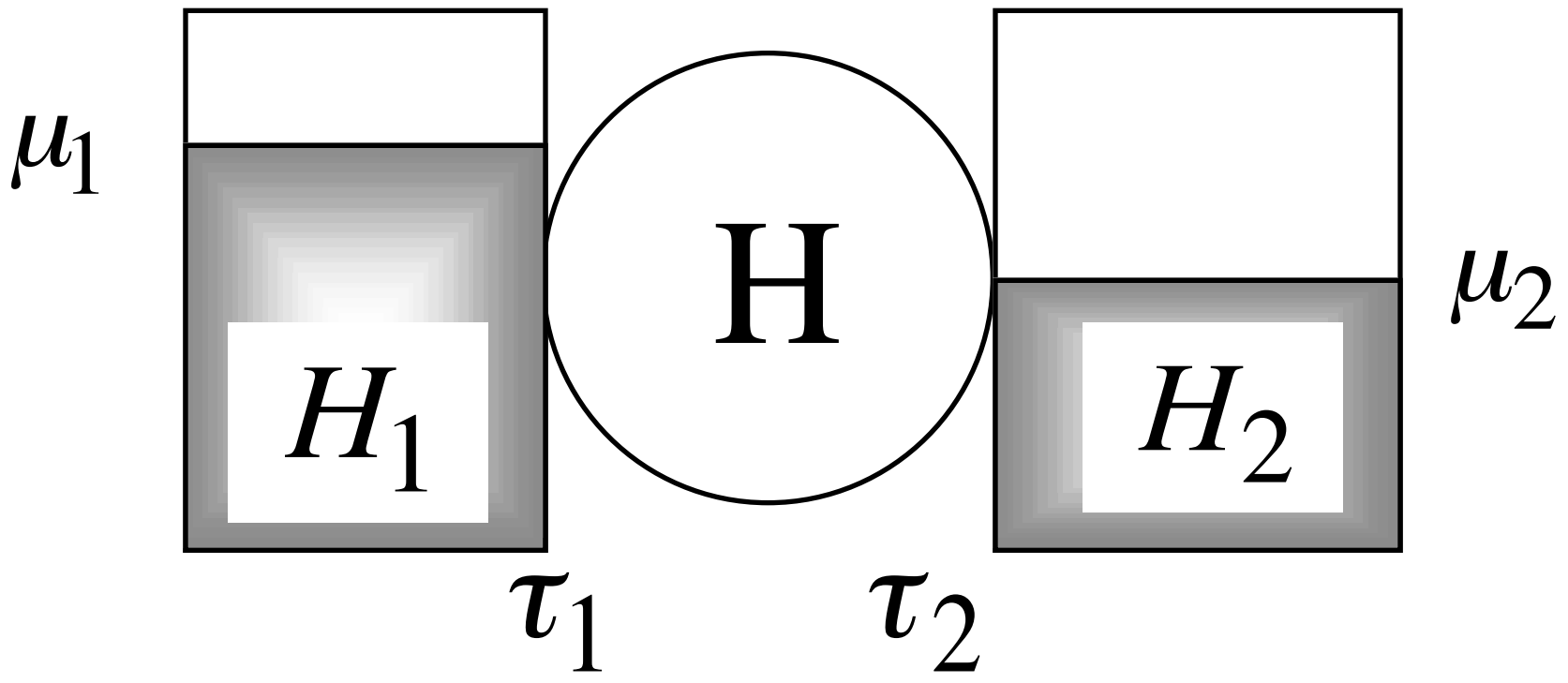

***Fig.2.1. Schematic representation of a strongly interacting system described by H coupled through electron exchange processes $\tau$ to multiple reservoirs or contacts; two are shown in the illustration.***

In the commonly used NEGF method, the interaction term in H (second term of Eq.(1.1)) is treated perturbatively through a self-energy function. However, this is inadequate for treating many problems of current interest where the interaction term is strong enough that a non-perturbative treatment is called for. A well-known example is the "Coulomb blockade" problem which is commonly addressed using the exact many-electron eigenstates of H (including the interaction term), with the contacts inducing transitions between states that differ by one electron. Our approach is formulated in this many-electron Fock space, but unlike the standard formulation, there is no need for the basis functions to be energy eigenstates since off-diagonal elements and energy broadening are included. We believe that the present approach can be extended to other kinds of strong interactions. For example, H could include strong electron-phonon interactions in addition to strong electron-electron interactions.

The system is described in terms of a many-electron basis $|I\rangle$ (not necessarily the eigenstates of H) while the contacts are described in terms of a basis $|R\rangle$ which are assumed to be the exact many-electron eigenstates of $H_1$ and $H_2$ with energies $\varepsilon_R$. We start from the equation for the time evolution of the composite wavefunction of the full channel + contact system**:**

$$i\hbar\frac{d}{dt}\tilde{\Psi}_{IR} - \sum_{I'} H_{II'}\,\tilde{\Psi}_{I'R} - \varepsilon_R\,\tilde{\Psi}_{IR} \quad = \quad \sum_{I',R'} [\tau]_{IR,I'R'}\,\tilde{\Psi}_{I'R'} \qquad (2.4)$$

and perform the transformation $\tilde{\Psi}_{IR}(t) = \tilde{\psi}_{IR}(t)\exp(-i\varepsilon_R t/\hbar)$ to obtain

$$i\hbar\frac{d}{dt}\tilde{\psi}_{IR} - \sum_{I'} H_{II'}\,\tilde{\psi}_{I'R} \quad = \quad \sum_{I',R'} [\tilde{\tau}(t)]_{IR,I'R'}\,\tilde{\psi}_{I'R'} \qquad (2.5)$$

$$\text{with} \quad [\tilde{\tau}(t)]_{IR,I'R'} \quad \equiv \quad [\tilde{\tau}]_{IR,I'R'}\exp(+i(\varepsilon_R - \varepsilon_{R'})t/\hbar) \qquad (2.6)$$

This is equivalent to treating the isolated contacts in the Heisenberg picture, so that they affect the operators, not the wavefunctions: however, the system and the interaction with the contacts are still in the Schrodinger picture.

The basic quantity in our approach is the many-electron wavefunction traced over the contact or reservoir degrees of freedom:

$$\rho_{IJ}(t,t') \quad \equiv \quad \sum_R \tilde{\psi}_{IR}(t)\,\tilde{\psi}_{JR}(t')^* \quad \equiv \quad \sum_R \{\psi_I(t)\}_R\,\{\psi_J(t')\}_R^+ \qquad (2.7)$$

The size of this many-electron density matrix depends on the number of many-electron states 'I' needed to describe the strongly interacting ***system***. For example, if the system has two one-electron energy levels, there are four multi-electron energy levels (see Fig.1.1). The density matrix $\rho_{IJ}$ is then a (4x4) matrix having non-zero off-diagonal elements between levels with the same number of electrons in normal systems.

The formulation presented here provides a compact set of equations that can be solved to obtain the Fock space density matrix knowing the correlation properties of the contacts. It is not common to calculate this Fock space density matrix, possibly because it is normally too large. But in nanoscale systems, this matrix can be small enough to be amenable at least to numerical computation and any quantity of interest can then be obtained from it.

## 2.2. "FSGF" Equations : Time-dependent version

In this section we summarize the general time-dependent equations which are derived in the ***Appendix***. Readers familiar with the NEGF formalism will recognize the similarity with the equations summarized below. The differences are discussed briefly in Section 2.4. Here we would like to note one basic distinction: All quantities appearing below like the Green function, self-energy, density matrix are defined in Fock space and as such are matrices of dimension ($2^N \text{x} 2^N$), N being the number of basis functions describing the one-electron Hilbert space. Similar quantities appearing in the NEGF formalism are of dimension (NxN).

The Fock space Green's function ("FSGF") equations can be grouped in three separate categories that are solved sequentially. The first set gives us the availability of states at different energies (after coupling to the contacts) through the Fock space Green function G; the second set gives us the occupation of these states through the Fock space density matrix $\rho$; the third set allows us to calculate different observables of interest, from a knowledge of G, $\rho$.

***2.2.1. Equations for Fock space Green function:*** These equations allow us to calculate the Green function, given (1) a system described by the many-electron Hamiltonian H (when isolated) and (2) a function F describing the correlation properties of the contacts (C).

$$i\hbar\frac{d}{dt}G_{IJ}(t,t') - \sum_{M} H_{IM}\, G_{MJ}(t,t') - \int_{-\infty}^{+\infty} dt_1 \sum_{M} \Sigma_{IM}(t,t_1)\, G_{MJ}(t_1,t') = \delta_{IJ}\,\delta(t-t') \tag{2.8}$$

$$\Sigma_{MN}(t,t') = \sum_{K,L} F_{MN,KL}(t,t')\, G_{KL}(t,t') \tag{2.9}$$

$$F_{MN,KL}(t,t') = \sum_{C} \left\langle [\tau(t)]_{MK}\, [\tau(t')]_{LN} \right\rangle_C \tag{2.10}$$

***2.2.2. Equations for Fock space density matrix:*** These equations allow us to calculate the density matrix, given (1) a system Green function G (after connecting to contacts) obtained from Eqs.(2.8)-(2.10) and (2) a function f describing the correlation properties of the contacts (C).

$$\rho_{IJ}(t,t') = \sum_{N,M} \int_{-\infty}^{+\infty} dt_1 \int_{-\infty}^{+\infty} dt_1'\; G_{IM}(t,t_1)\, S_{MN}(t_1,t_1')\, G_{JN}(t',t_1')^* \tag{2.11}$$

$$S_{MN}(t,t') = \sum_{K,L} f_{MN,KL}(t,t')\, \rho_{KL}(t,t') \tag{2.12}$$

$$f_{MN,KL}(t,t') = \sum_{C} \left\langle [\tau(t')]_{LN}\, [\tau(t)]_{MK} \right\rangle_C = F_{LK,NM}(t_1',t_1) \tag{2.13}$$

***2.2.3. Equations for calculating observables:*** From a knowledge of the density matrix $\rho$, these equations allow one to calculate the expectation value of any observable Q described by operator $\Theta_Q$

$$\langle Q(t)\rangle = Trace\,[\rho(t,t)]\,[\Theta_Q] \tag{2.14}$$

$$\frac{d}{dt}\langle Q(t)\rangle = \frac{1}{i\hbar} Trace \begin{pmatrix} [\rho(t,t)][\Theta_Q H - H\Theta_Q] \\ +\int dt_1\, [\Sigma(t,t_1)][\rho(t_1,t)]\Theta_Q - [\rho(t,t_1)][\Sigma^+(t_1,t)]\Theta_Q \\ +\int dt_1\, [S(t,t_1)][G^+(t_1,t)]\Theta_Q - [G(t,t_1)][S(t_1,t)]\Theta_Q \end{pmatrix} \tag{2.15}$$

and the correlation of two variables P and Q ( operators $\Theta_P$ and $\Theta_Q$ respectively)

$$\langle P(t)Q(t')\rangle = Trace\,[\rho(t',t)]\,[\Theta_P]\,[U(t,t')]\,[\Theta_Q]$$

$$\text{where}\quad U(t,t') = i[G(t,t') - G^+(t,t')] \tag{2.16}$$

### 2.3. "FSGF" equations: Steady-state version

The equations we actually use for our examples in Section 3, are the steady-state versions of those listed in Section 2.2 (and derived in the Appendix) obtained by assuming that all quantities like say $\rho_{IJ}(t,t')$ in Section 2.2 are independent of (t + $t'$), and Fourier transforming with respect to (t - $t'$) --> E,

$$\rho_{IJ}(t,t') \rightarrow \rho_{IJ}(t-t') = \int_{-\infty}^{+\infty} dE\, \rho_{IJ}(E)\, \exp(-iE(t-t')/\hbar) \tag{2.17}$$

This leads to the following steady-state equations corresponding to the time-dependent versions in Section 2.2.

#### *2.3.1. Equations for Fock space Green function :*

$$E\, G_{IJ}(E) - \sum_J H_{IM}\, G_{MJ}(E) - \sum_M \Sigma_{IM}(E)\, G_{MJ}(E) = \delta_{IJ}$$

$$\rightarrow [EI - H - \Sigma(E)]G(E) = I \tag{2.18}$$

$$\Sigma_{MN}(E) = \int_{-\infty}^{+\infty} d\varepsilon\, F_{MN,KL}(\varepsilon)\, G_{KL}(E-\varepsilon) \tag{2.19}$$

#### *2.3.2. Equations for Fock space density matrix:*

$$\rho_{IJ}(E) = \sum_{N,M} G_{IM}(E)\, S_{MN}(E)\, [G^+]_{NJ}(E)$$

$$\rightarrow [\rho(E)] = [G(E)]\,[S(E)]\,[G^+(E)] \tag{2.20}$$

$$S_{MN}(E) = \int_{-\infty}^{+\infty} d\varepsilon\, f_{MN,KL}(\varepsilon)\, \rho_{KL}(E-\varepsilon) \tag{2.21}$$

#### *2.3.3. Equations for calculating observables:*

$$\langle Q \rangle = \int_{-\infty}^{+\infty} dE\; Trace\, [\rho(E)]\,[\,\Theta_Q] \tag{2.22}$$

$$\langle I_Q \rangle = \frac{i}{\hbar} \int_{-\infty}^{+\infty} dE\; Trace \left\{ \begin{pmatrix} [H\rho(E) - \rho(E)H] \\ [G(E)]\,[S(E)] - [S(E)][G^+(E)] \\ +[\rho(E)][\Sigma^+(E)] - [\Sigma(E)][\rho(E)] \end{pmatrix} [\Theta_Q] \right\} \tag{2.23}$$

$$\overline{PQ}(\varepsilon) = \int_{-\infty}^{+\infty} dE\; Trace\, [\rho(E)]\,[\Theta_P]\,[U(E-\varepsilon)]\,[\Theta_Q]$$

$$\text{where} \quad U(E) = i[G(E) - G^+(E)] \tag{2.24}$$

***2.3.4. Current calculation:*** Typically we are interested in the terminal current, which can be obtained from eq.(2.15) or (2.23), by choosing the quantity 'Q" as the charge, so that $\Theta_Q$ is equal to (- e) times the number operator. However, if we evaluate eq.(2.23) directly, the answer will be zero, because it gives the net current flowing into all the contacts which should indeed be zero under steady-state conditions. What we need is the current at each contact ***individually*** which should all add up to zero. The individual currents are obtained by evaluating the current expressions (eq.(2.15) or (2.23)) using only the components of the source function S and the self-energy function $\Sigma$ that arise from the 'f' (see eqs. (2.10) and (2.19)) or 'F' (see eqs.(2.13) and (2.21)) due to the contact of interest alone (and not summed over all contacts 'C' as indicated in eqs.(2.10) and (2.13)).

***2.3.5. Current conservation:*** As mentioned above, the total current into all the contacts should add up to zero at steady-state. It is shown in the Appendix that this is insured provided

$$F_{MNKL}(t,t') = f_{LKNM}(t',t) \tag{2.25a}$$

which is automatically true according to our prescription (see eq.(2.13)). Under steady-state conditions, Eq.(2.25a) can be Fourier transformed to yield

$$F_{MNKL}(\varepsilon) = f_{LKNM}(-\varepsilon) \tag{2.25b}$$

***2.4. NEGF versus "FSGF":***

As noted earlier the basic distinction between the standard NEGF equations and the "FSGF" equations proposed here is that all quantities in NEGF are of dimension (NxN), while in "FSGF" they are of dimension ($2^N$ x $2^N$), N being the number of one-electron basis functions. Two other important distinctions are noted below.

Note that in writing the equations for the functions $\Sigma$ and S describing the effect of contacts, we have used the equivalent of the "self-consistent Born approximation" often used in NEGF. Interestingly, when an electron of energy $\varepsilon$ enters from a contact, the system makes a transition in Fock space between states of energy E and $E+\varepsilon$, which is reminiscent of inelastic processes in one-electron space. "FSGF" equations even for elastic transport thus look similar to NEGF equations for inelastic transport. A second important distinction is that the "FSGF" equations for the Green function G and those for the density matrix $\rho$ are decoupled from each other and can be solved sequentially. By contrast, the NEGF equations for the Green function $G^R$ and for the correlation function $-iG^<$ are coupled and have to be solved iteratively. We believe this coupling arises from the exclusion principle and is absent in the present formulation because in Fock space there is no exclusion principle to worry about explicitly; it is already included in our choice of basis functions (Fig.1.1b). A system is never "Pauli blocked" from entering any state in Fock space.

Finally we note that once the Green's function G and density matrix $\rho$ in Fock space have been obtained, we can calculate any of the usual quantities in the NEGF formalism like the one-particle spectral function (A) or the correlation function ( $-iG^<$) by convolving different Fock space quantities as indicated in Eq.(2.24). In this paper we present only one example of a calculation of this type, namely the spectral function for a quantum dot in the Kondo regime (see Fig.4.1).

## 3. Application

As mentioned in the introduction we will illustrate our formalism by describing its application to the structure shown in Fig.1.3. with four spin-polarized contacts. This requires a numerical solution of the "FSGF" equations involving three sequential steps:

(1) Solve eqs.(2.18) and (2.19) self-consistently for the functions G, Σ, using the Hamiltonian [H] and the contact coupling function F.

(2) Solve eqs.(2.20) and (2.21) self-consistently using the Green function G from step 1 and the contact coupling function f to obtain $\rho$, S.

(3) Evaluate the current from eq.(2.23) using G, Σ from step 1 and $\rho$, S from step 2.

In Section 3.1, we will describe how the input Hamiltonian H and contact correlation functions F and f are set up. We will then present (Section 3.2) some numerical results obtained from our formalism showing evidence for spin-blockade and negative differential resistance (NDR) in the I-V characteristics when the contacts are polarized along z (see Fig.1.3, with $\theta$ = 0, Section 3.2). Finally in Section 3.3 we will describe how the spin-blockade is modified when the contact polarization is along x (see Fig.1.3, with $\theta = \pi/2$).

## 3.1. Model inputs

***3.1.1. Hamiltonian [H]:*** The system Hamiltonian [H] in Fock space is a (4x4) matrix:

$$\mathbf{H} = \begin{array}{c|cccc} & 0 & 1 & 2 & 3 \\ \hline 0 & 0 & 0 & 0 & 0 \\ 1 & 0 & \varepsilon_\downarrow & 0 & 0 \\ 2 & 0 & 0 & \varepsilon_\uparrow & 0 \\ 3 & 0 & 0 & 0 & \varepsilon_\downarrow + \varepsilon_\uparrow + U_0 \end{array} \qquad (3.1)$$

where $\varepsilon_\uparrow$ and $\varepsilon_\downarrow$ are the energies of a non-interacting spin-up and a non-interacting spin-down electron respectively and we have used 0, 1, 2 and 3 to denote the basis functions (00), (01), (10) and (11) in Fig.1.1b.

***3.1.2. Contact correlation functions (f, F):*** To calculate these functions (see eqs.(2.10) and (2.13)) we need the coupling matrix for each contact C, which has the form

$$\boldsymbol{\tau} = \begin{array}{c|cccc} & 0 & 1 & 2 & 3 \\ \hline 0 & 0 & \tau_d^{+} & \tau_u^{+} & 0 \\ 1 & \tau_d & 0 & 0 & \tau_u^{+} \\ 2 & \tau_u & 0 & 0 & \tau_d^{+} \\ 3 & 0 & \tau_u & \tau_d & 0 \end{array} \qquad (3.2)$$

It is apparent from eqs.(2.10) and (2.13) that the function F(t, t') involves quantities like

$$\left\langle \tau_i^{+}(t)\,\tau_j(t') \right\rangle \qquad \text{or} \qquad \left\langle \tau_i(t)\,\tau_j^{+}(t') \right\rangle$$

where the indices 'i' and 'j' stand for either up ('u') or down ('d'), while the function f(t,t') involves the time-reversed versions of these quantities with t and t' interchanged. For any contact C, we can write from eq.(2.3),

$$\tau_u = \sum_{r \in C} t_{ur} c_r \quad , \qquad \tau_d = \sum_{s \in C} t_{ds} c_s \tag{3.3}$$

Consider a specific element of F, say $F_{00;21}$. We can write

$$F_{00;21}(t,t') = \left\langle [\tau_u^+(t)][\tau_d(t')] \right\rangle = \sum_{r,s \in C} t_{u,r}{}^* t_{d,s} \left\langle [c_r^+(t)][c_s(t')] \right\rangle$$
$$= \sum_{r \in C} |t(\varepsilon_r)|^2 (-c^* s)\, f_0(\varepsilon_r - \mu_C) \exp[+i\varepsilon_r (t-t')] \tag{3.4}$$

where $f_0(\varepsilon - \mu)$ represents the Fermi function with a electrochemical potential $\mu$ and it is assumed that the up and down states in the device couple with the corresponding spin states in the contact with identical matrix elements $t_C$ for all contact states around the same energy. The factor c*s arises from the fact that the eigenstates of the contact are polarized at an angle $(\theta, \phi)$ with respect to the up ('u') and down ('d') spin states along 'z' (see Fig.1.3):

$$c \equiv \cos(\theta_C/2)\, e^{+i\phi_C/2} \qquad s \equiv \sin(\theta_C/2)\, e^{-i\phi_C/2} \tag{3.5}$$

For a down-polarized contact, $c \to -s^*$ and $s \to c^*$. Fourier transforming eq.(3.4) we obtain,

$$F_{00;21}(\varepsilon) = \sum_{r \in C} |t(\varepsilon_r)|^2 (-c^* s)\, f_0(\varepsilon - \mu_C)\, 2\pi\, \delta(\varepsilon + \varepsilon_r) \tag{3.6}$$

Replacing the summation with an integral over the contact density of states $D_C(\varepsilon_R)$

$$F_{00;21}(\varepsilon) = \int_{-\infty}^{+\infty} d\varepsilon_r\, D_C(\varepsilon_r) |t(\varepsilon_r)|^2 (-c^* s)\, f_0(\varepsilon_r - \mu_C)\, 2\pi\, \delta(\varepsilon + \varepsilon_r)$$
$$= (-c^* s)\; \gamma_C\, f_0(-\varepsilon - \mu_C)$$

where $\gamma_C \equiv 2\pi D_C(\varepsilon) |t(\varepsilon)|^2$ (3.7)

This is of course the standard golden rule expression for the contact coupling; for our examples we will treat $\gamma_C$ for the different contacts as input parameters.

Proceeding similarly we can evaluate all the relevant elements of the functions F which can be expressed in terms of the following functions:

$$F_R(\varepsilon) = \gamma_C f_0(-\varepsilon + \mu_C) \qquad F_A(\varepsilon) = \gamma_C f_0(-\varepsilon - \mu_C) \tag{3.8a}$$

and are tabulated here for easy reference:

| $F$ $\downarrow MN$, $KL \rightarrow$ | 00 | 11 | 22 | 33 | 12 | 21 |
|---|---|---|---|---|---|---|
| 00 | 0 | $F_A s^* s$ | $F_A c^* c$ | 0 | $-F_A s^* c$ | $-F_A c^* s$ |
| 11 | $F_R s^* s$ | 0 | 0 | $F_A c^* c$ | 0 | 0 |
| 22 | $F_R c^* c$ | 0 | 0 | $F_A s^* s$ | 0 | 0 |
| 33 | 0 | $F_R c^* c$ | $F_R c^* c$ | 0 | $F_R s^* c$ | $F_R c^* s$ |
| 12 | $-F_R s^* c$ | 0 | 0 | $F_A s^* c$ | 0 | 0 |
| 21 | $-F_R c^* s$ | 0 | 0 | $F_A c^* s$ | 0 | 0 |

(3.8b)

Note that off-diagonal combinations of KL or MN are needed for states 1 and 2, since only these have the same number of electrons. Combinations like 03 or 30 involving states with number of electrons differing by an even number of electrons are possible in principle (as in the superconducting state) and could be included within our framework, but we do not consider them in this paper.

Similarly the relevant elements of f can be expressed in terms of

$$f_R(\varepsilon) = \gamma_C f_0(\varepsilon + \mu_C) \qquad f_A(\varepsilon) = \gamma_C f_0(\varepsilon - \mu_C) \tag{3.9a}$$

$$
\begin{array}{lcccccc}
f & KL \rightarrow & & & & & \\
\downarrow MN & 00 & 11 & 22 & 33 & 12 & 21 \\
00 & 0 & f_R\, s^*s & f_R\, c^*c & 0 & -f_R s^*c & -f_R c^*s \\
11 & f_A\, s^*s & 0 & 0 & f_R\, c^*c & 0 & 0 \\
22 & f_A\, c^*c & 0 & 0 & f_R s^*s & 0 & 0 \\
33 & 0 & f_A\, c^*c & f_A\, c^*c & 0 & f_A s^*c & f_A c^*s \\
12 & -f_A\, s^*c & 0 & 0 & f_R s^*c & 0 & 0 \\
21 & -f_A c^*s & 0 & 0 & f_R c^*s & 0 & 0
\end{array}
\qquad (3.9b)
$$

Eqs.(3.8a, b) and (3.9a, b) can be used to obtain the contact coupling functions for each contact ‘C’. In addition we include a self-scattering process that is assumed to scatter each state back into itself, leading to non-zero contact coupling functions of the form:

$$F_{MM;MM} = \gamma_B\,, \qquad f_{MM;MM} = \gamma_B \qquad (3.10)$$

This additional broadening $\gamma_B$ facilitates numerical computations by making it feasible to work with relatively coarser energy grids, but it does not seem to affect the physics in any significant way. For example, the widths of the nulls in Fig.3.4 is much less than the value of $\gamma_B$ used in these calculations, 10 meV and is determined solely by the contact broadening.

### 3.2. Spin blockade with one z-polarized contact

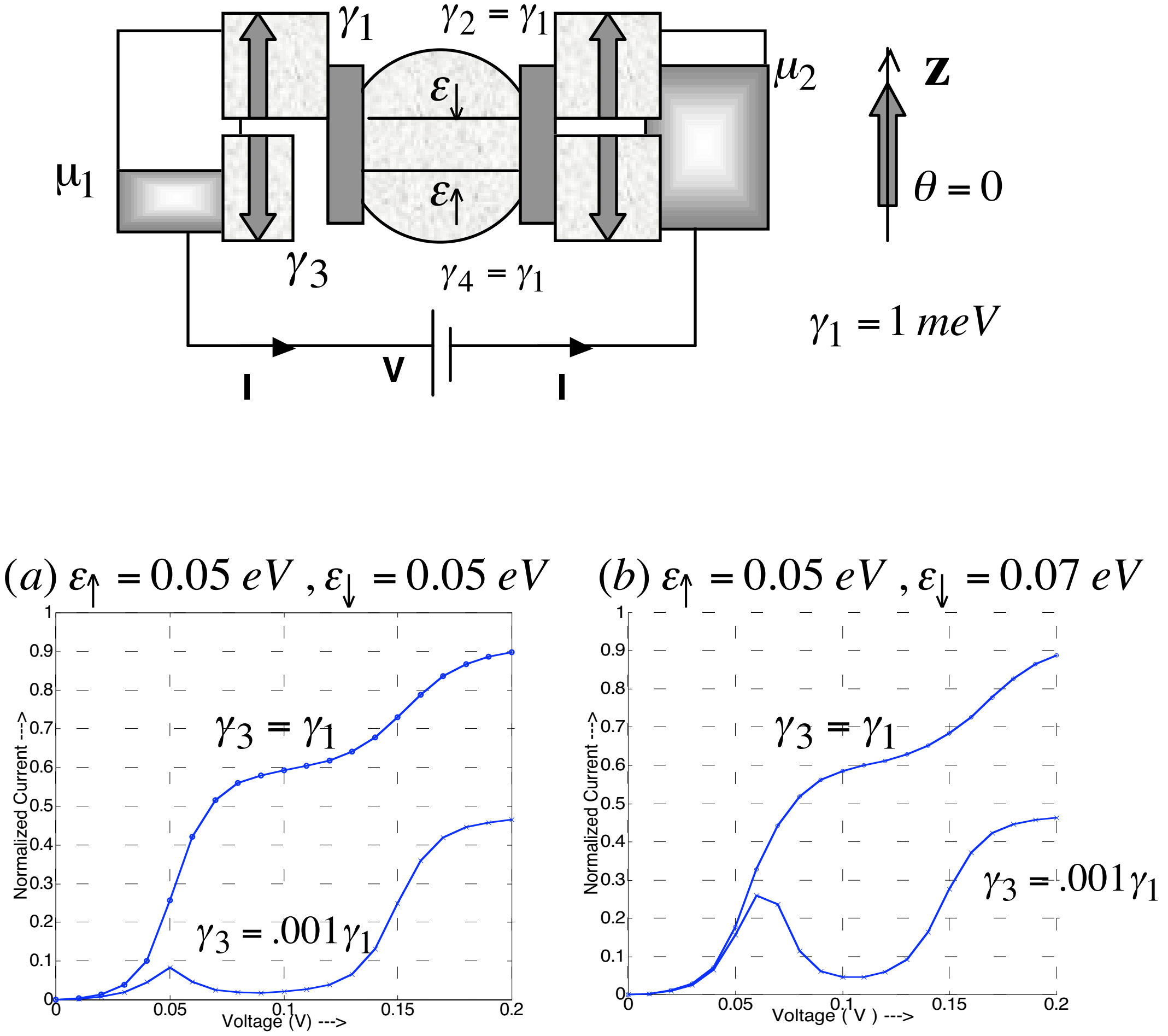

***Fig.3.1. Current versus voltage with the left contact unpolarized (** $\gamma_3 = \gamma_1$**) and z-polarized (** $\gamma_3$**=0.001** $\gamma_1$**). The latter shows negative differential resistance (NDR) due to "spin-blockade" which is enhanced if** $\varepsilon_\downarrow$ **is raised above** $\varepsilon_\uparrow$**. Current is normalized to*** $(e/\hbar)2\gamma_1\gamma_2/(\gamma_1+\gamma_2)$.

We will now describe some numerical results obtained by solving the steady-state equations described in Section 2.3. We consider a structure with one up-spin and one down-spin level connected to four contacts as shown in Fig.3.1. The right contact is unpolarized, that is, both up and down-spin components are equally coupled to the device: $\gamma_4 = \gamma_2 = 1$ meV. If the left contact too is unpolarized $\gamma_3 = \gamma_1 = 1$ meV, then we obtain a broadened Coulomb blockade curve like the one shown earlier in Fig.1.2. But if the left contact is polarized with the down component literally disconnected from the device $\gamma_3 = 0.001\,\gamma_1 = 1\,\mu$eV, we obtain a different I-V characteristic showing evidence for "spin-blockade": Current starts to flow through the up-spin level but quickly shuts off when the down-spin level fills up and forces the up-spin level to float up in energy. The resulting negative differential resistance (NDR) is enhanced if $\varepsilon_\downarrow$ is raised above $\varepsilon_\uparrow$ so that the down-spin level starts to fill up a little later. This is evident from Figs.3.2a, b showing the occupation probabilities for up- and down-spin states.

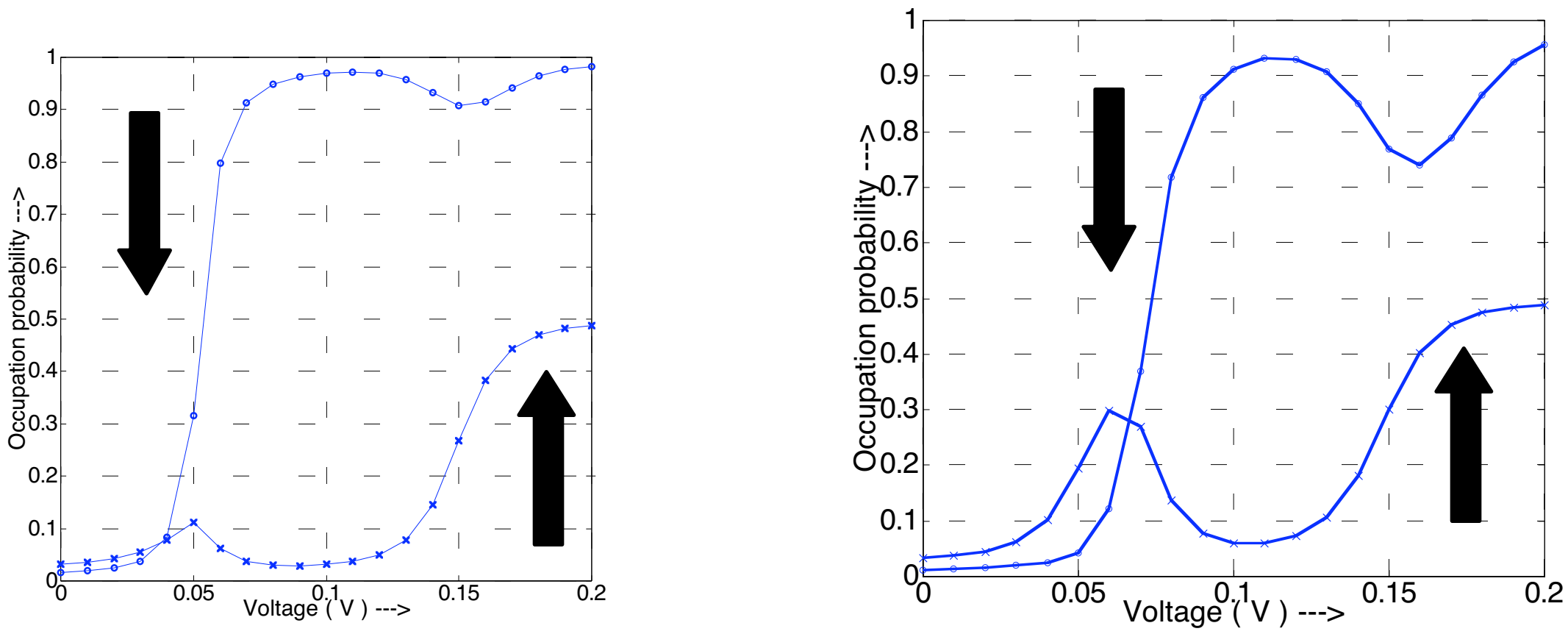

***Fig.3.2. Occupation probabilities of up- and down- spin states corresponding to Fig.3.1a, b. Note how the increase in down-spin occupation shuts off the rise in the up-spin occupation.***

## 3.3. Lifting of spin blockade with one x-polarized contact

The example shown in Figs.3.1 and 3.2 requires both Coulomb blockade and broadening but not quantum interference. The physics is described by the diagonal elements of the density matrix and the off-diagonal elements $\rho_{12}$, $\rho_{21}$ are zero. However, if we choose contacts that are polarized at 90 degrees to the z-direction (Fig.3.3) then ***the off-diagonal elements*** $\rho_{12}$, $\rho_{21}$ ***are essential to the physics*** since left- and right- polarized spins are linear combinations of up- and down- spins. If the up-spin and down-spin levels are non-degenerate (Figs.3.1b and 3.3b) then the I-V characteristics are very different: there is no spin-blockade when the contact spins are turned sideways (Fig.3.3b).

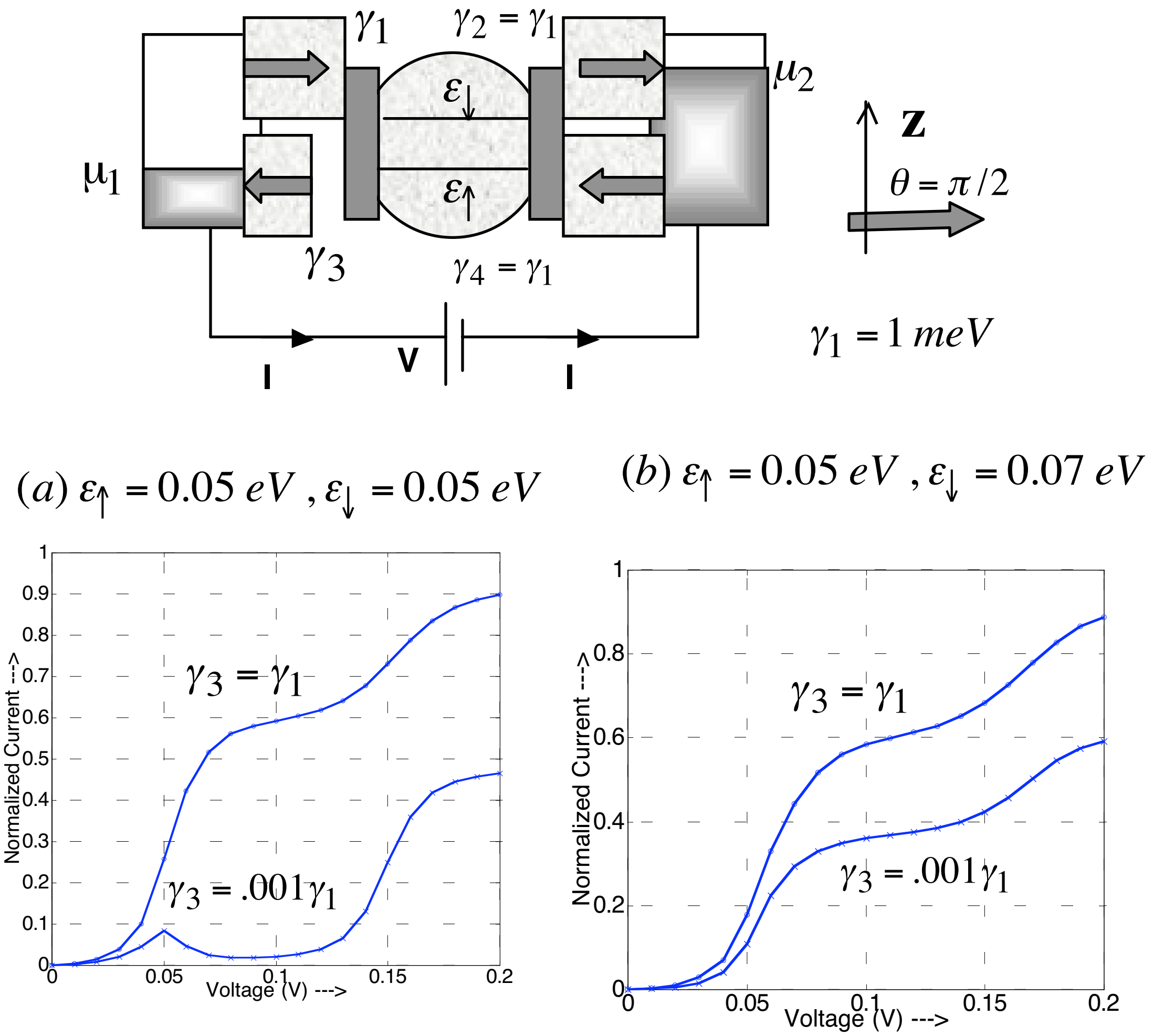

***Fig.3.3. Current versus voltage with the left contact unpolarized (*** $\gamma_3 = \gamma_1$***) and x-polarized (*** $\gamma_3 = 0.001\,\gamma_1$***). The latter shows negative differential resistance (NDR) due to "spin-blockade" if (a)*** $\varepsilon_\downarrow = \varepsilon_\uparrow$ ***but not if (b) the spin-degeneracy is lifted:*** $\varepsilon_\downarrow \neq \varepsilon_\uparrow$***. Current is normalized to*** $(e/\hbar)2\gamma_1\gamma_2/(\gamma_1+\gamma_2)$***.***

However, if the up- and down-spins are degenerate, there is no change in the I-V characteristics (Fis.3.1a, 3.3a) when the contact polarization is turned by 90 degrees. This is to be expected since the physics of degenerate levels could just as well been described using left and right- spins as the basis (rather than up- and down) and the structure in Fig.3.3 would then be identical to Fig.3.1 with the spins turned sideways. The fact that the I-V characteristics remain unchanged, however, provides a good test for our formalism since in one case (Fig.3.1a) the off-diagonal elements $\rho_{12}$, $\rho_{21}$ are zero, while in the other case (Fig.3.3a) they are comparable in magnitude to the diagonal elements $\rho_{11}$ and $\rho_{22}$. ***This basis –independence is a crucial test of any quantum transport formalism.***

Fig.3.4 shows the variation in the current as the degeneracy $(\varepsilon_\uparrow - \varepsilon_\downarrow)$ of the two levels is tuned through zero (keeping the lower level at 0.05 eV), at a fixed voltage (0.1 V) which is chosen such that one would expect current suppression by spin blockade for degenerate levels with $\varepsilon_\uparrow - \varepsilon_\downarrow = 0$. Note that the current is virtually unchanged for z-polarized contacts, but shows a sharp increase for contacts polarized at 90 degrees to z. The width of the "null" is ~ $\gamma$, the broadening introduced by the coupling to the contacts.

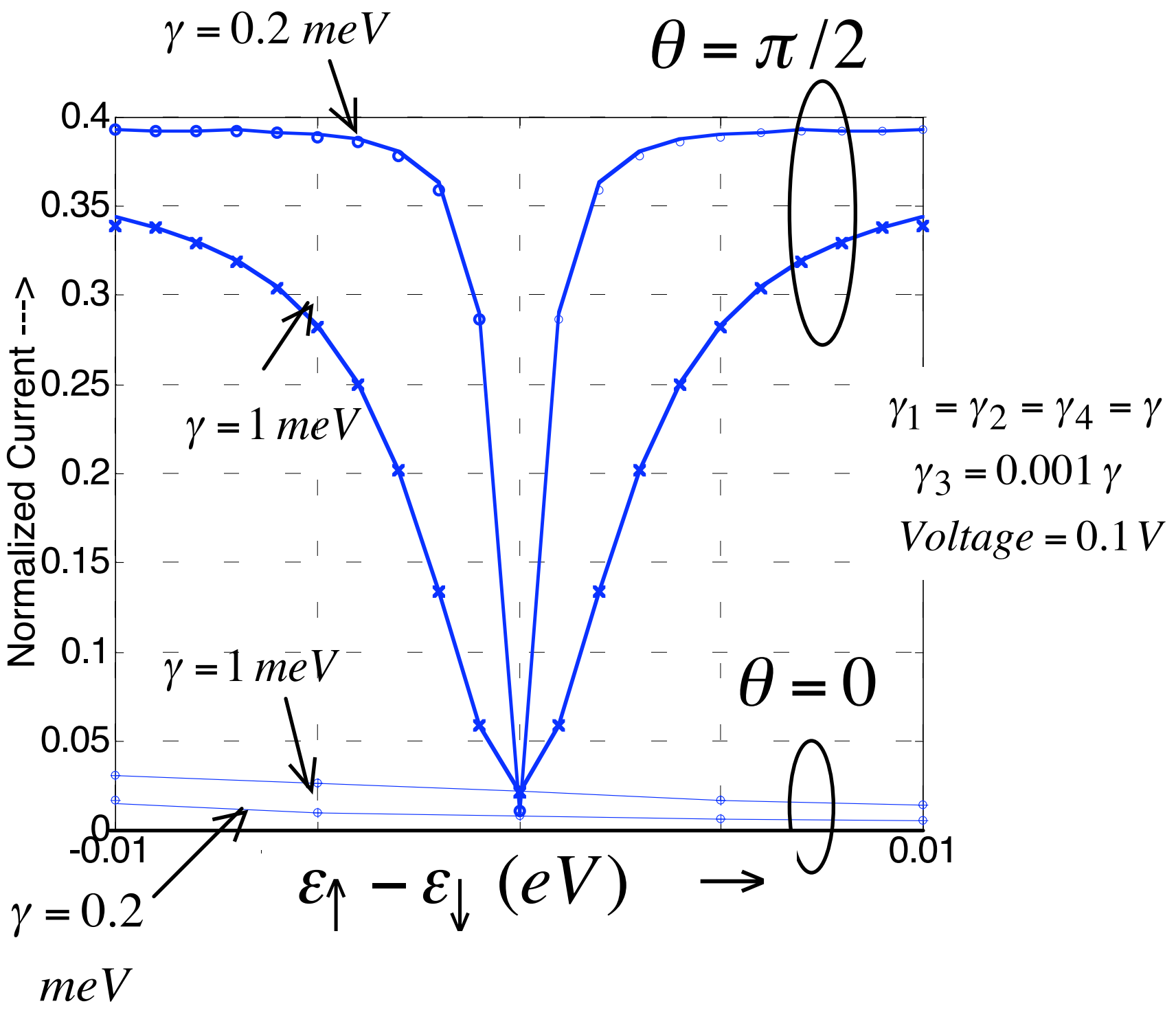

***Fig.3.4. Current at a voltage of 0.1 V as a function of the spin degeneracy :*** $\varepsilon_\uparrow - \varepsilon_\downarrow$***. Note how the spin blockade remains unaffected if the left contact is z-polarized (*** $\theta$***= 0), but is lifted if the left contact is x-polarized (*** $\theta = \pi/2$***). In the latter case the width of the null is determined by the broadening due to the coupling to the contacts. Current is normalized to*** $(e/\hbar)2\gamma_1\gamma_2/(\gamma_1+\gamma_2)$.

## 4. Concluding remarks

In summary, we have presented a Fock space formulation for nanoscale electronic transport and illustrated it with an example that demonstrates all three effects that we are trying to capture: strong interaction (large $U_0$) leading to Coulomb or spin blockade, broadening $\gamma$ due to coupling to contacts and quantum coherence effects due to spin polarization in directions other than the z-direction. The standard master equation method for Coulomb blockade captures the first effect, while standard NEGF methods capture the last two. But there is a need for a method that captures all three and in this paper we have proposed an approach using "Fock space Green's functions" (FSGF) that appears to fulfill this need. Although the contact is treated only within the "self-consistent Born approximation", we find that the one-electron spectral function (calculated by convolving Fock space Green functions (FSGF), see Eq.(2.24)) shows evidence for Kondo peaks in the appropriate range of parameters (see Fig.4.1). The basic equations for the FSGF approach are quite general and we leave it to future work to assess how effective it will be for more interesting and exotic transport problems involving the entanglement of many-electron states.

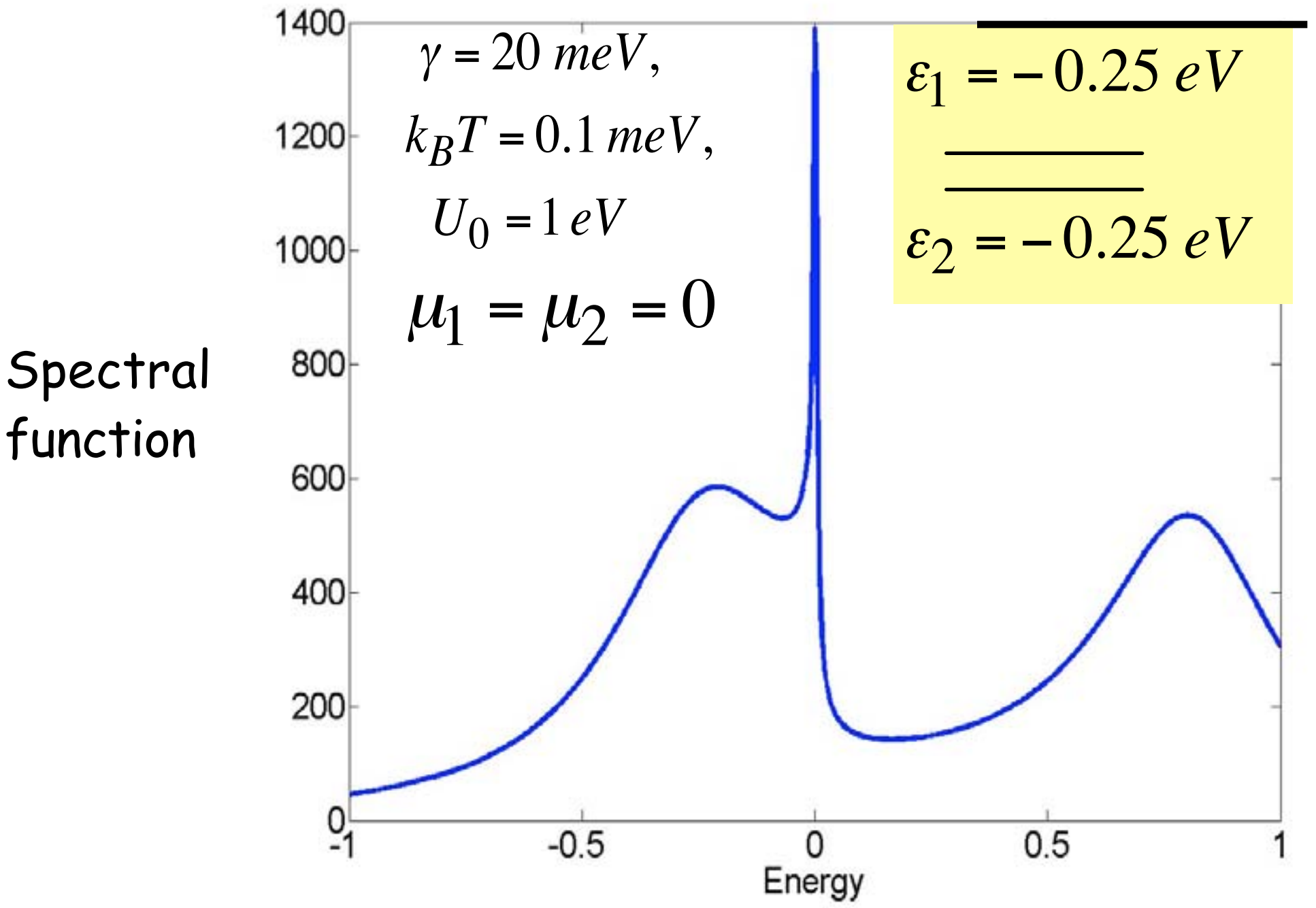

***Fig.4.1. Spectral function calculated by convolving Fock space Green functions (see Eq.(2.24)) showing evidence for Kondo peaks similar to those presented in Ref. 10.***

**Acknowledgements**
I am indebted to Bhaskaran Muralidharan for many helpful suggestions throughout the preparation of this manuscript and for providing me with Fig.4.1. I am also grateful to Diptiman Sen and Roger Lake for their suggestions. This work was supported by the National Science Foundation through the Network for Computational Nanotechnology.

**Appendix A: Derivation of Eqs.(2.8)-(2.16), (2.25)**

Our objective in this section is to provide a derivation of the full time-dependent equations summarized in Section 2.2 (see eqs.(2.8) through (2.16)), along with the condition for conservation of probability stated in eq.(2.25).

***A.1. Equations for the Fock space Green's function (eqs.(2.8), (2.9) and (2.10)):***

We start from eq.(2.5) which can be written in matrix form as

$$\left( i\hbar \frac{d}{dt} - [\tilde{H}] - [\tilde{\tau}(t)] \right) \{\tilde{\psi}\} \quad = \quad \{0\} \tag{A.1}$$

where $\tilde{H}_{IR,I'R'} \;=\; H_{I,I'}\,\delta_{R,R'}$. We can define a Green's function for eq.(A.1) as

$$\left( i\hbar \frac{d}{dt} - [\tilde{H}] - [\tilde{\tau}(t)] \right) [\tilde{G}(t,t')] \quad = \quad [\tilde{I}]\,\delta(t-t') \tag{A.2}$$

We will generally use symbols with a tilde (~) to denote quantities that refer to the full channel (I) + contact (R) system, while we will use symbols without the tilde to denote channel quantities like the reduced Green's function obtained by averaging over the contact (or reservoir) states:

$$G_{IJ}(t,t') \;\equiv\; \sum_R P_R\, \tilde{G}_{IR,JR}(t,t') \;\rightarrow\; \sum_R P_R\, [G_{IJ}(t,t')]_{RR}$$

that is, $\qquad [G(t,t')] \;\equiv\; Trace_R\,[\rho_R\,\tilde{G}(t,t')] \qquad$ (A.3)

where $\rho_R$ is the contact density matrix assumed to be approximately diagonal: $P_R\,\delta_{R,R'}$. We can obtain an equation for the reduced Green's function

$$i\hbar \frac{d}{dt} G_{IJ}(t,t') - \sum_M H_{IM}\, G_{MJ}(t,t') - \int_{-\infty}^{+\infty} dt_1 \sum_M \Sigma_{IM}(t,t_1)\; G_{MJ}(t_1,t') \;=\; \delta_{IJ}\,\delta(t-t') \qquad \text{(A.4, same as eq.(2.8))}$$

using methods similar to those used in many-body physics as outlined below. Note, however, that these methods are not commonly used in Fock space in the way we are using it. Indeed we are not aware of any other treatment along these lines.

We first note that the full Green's function defined in Eq.(A.2) obeys the integral equation:

$$[G(t,t')]_{RR} = [g(t,t')] + \int_{-\infty}^{+\infty} dt_1 \sum_{R'} [g(t,t_1)][\tau(t_1)]_{RR'}[G(t_1,t')]_{R'R} \qquad \text{(A.5)}$$

where $[g(t,t')]$ represents the Green's function for the isolated channel:

$$\left(i\hbar\frac{d}{dt} - [H]\right)[g(t,t')] = [I]\,\delta(t-t') \qquad \text{(A.6)}$$

From Eq.(A.5) we can write a solution for the full Green's function $[G(t,t_1)]_{RR}$ in the form of a series whose terms can be represented diagrammatically as

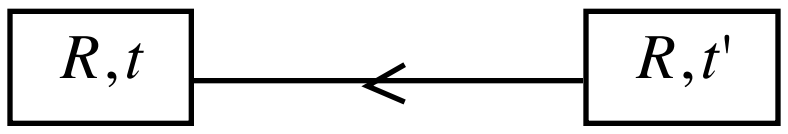

$$g(t,t')$$

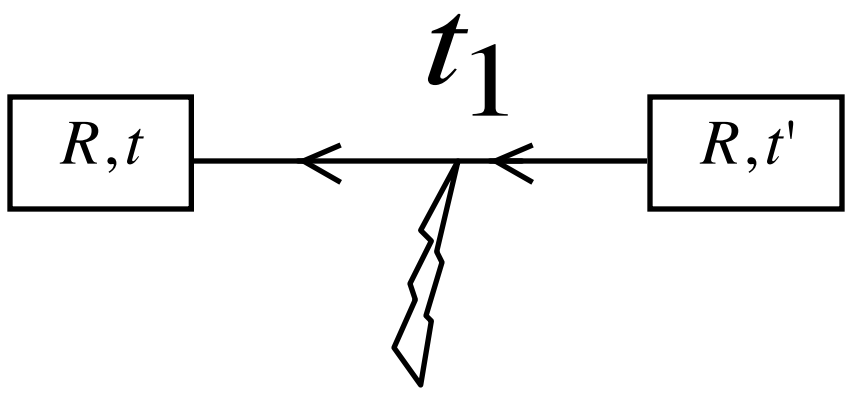

$$\int dt_1\, [g(t,t_1)][\tau(t_1)]_{RR}[g(t_1,t')]$$

(A.7)

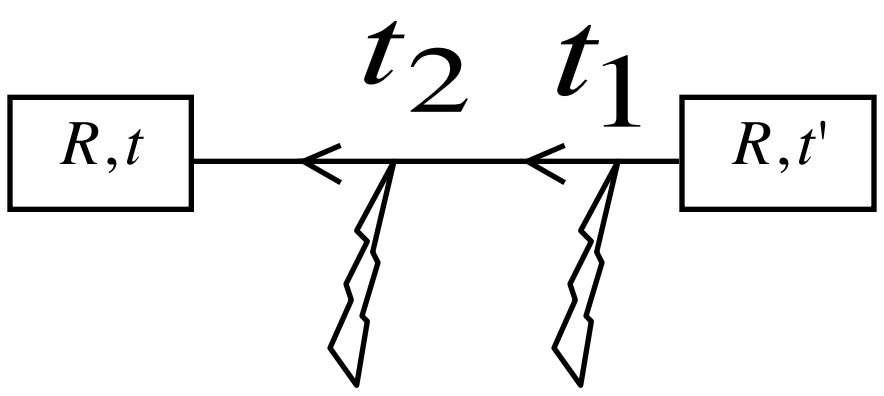

$$\int dt_1 \int dt_2\, [g(t,t_2)][\tau(t_2)]_{RR'}[g(t_2,t_1)][\tau(t_1)]_{R'R}[g(t_1,t')]$$

and so on. Since the elements of $[\tau(t)]_{RR'}$ are non-zero only for $R' \neq R$, it can be shown that the only terms of this series that are non-zero are those that contain an even number of interaction lines which are "tied together" in pairs: for every interaction going from one reservoir state to another, there is one bringing it back to the same state. One can then write a ***Dyson equation*** for the reduced Green's function

$$[G(t,t')]_{RR} = [g(t,t')] + \int_{-\infty}^{+\infty} dt_1 \int_{-\infty}^{+\infty} dt_2 \, [g(t,t_2)][\Sigma(t_2,t_1)]_{RR} [G(t_1,t')]_{RR}$$

in terms of the self-energy function $[\Sigma]_{RR}$ which can be calculated to any desired degree of accuracy by summing a suitable subset of irreducible diagrams. Neglecting any correlation between $\Sigma_{RR}$ and $G_{RR}$, we can average over the contact states 'R' to write

$$[G(t,t')] = [g(t,t')] + \int_{-\infty}^{+\infty} dt_1 \int_{-\infty}^{+\infty} dt_2 \, [g(t,t_2)][\Sigma(t_2,t_1)][G(t_1,t')] \qquad \text{(A.8)}$$

from which we obtain eq.(A.4) for the Green's function stated earlier by operating from the left with $\left(i\hbar \, d/dt - [H]\right)$ and making use of eq.(A.6).

The self-energy could be evaluated to different degrees of approximation. In this paper we use what corresponds to the ***"self-consistent Born approximation"***

$$[\Sigma_{MN}(t,t')]_{RR} = \sum_{R',K,L} \left[\tau(t)\right]_{MR,KR'} G_{KL}(t,t') \left[\tau(t')\right]_{LR',NR} \qquad \text{(A.9)}$$

so that

$$\Sigma_{MN}(t,t') = \sum_{R,R',K,L} P_R \left[\tau(t)\right]_{MR,KR'} G_{KL}(t,t') \left[\tau(t')\right]_{LR',NR}$$

$$= \sum_{K,L} F_{MNKL}(t,t') \, G_{KL}(t,t') \qquad \text{(A.10, same as eq.(2.9))}$$

where

$$F_{MNKL}(t,t') = \sum_{R,R'} P_R \left[\tau(t)\right]_{MR,KR'} \left[\tau(t')\right]_{LR',NR}$$

$$= Trace_R \left[\rho_R\right]\left[\tau(t)\right]_{MK} \left[\tau(t')\right]_{LN}$$

$$\equiv \left\langle \left[\tau(t)\right]_{MK} \left[\tau(t')\right]_{LN} \right\rangle_R \qquad \text{(A.11, same as eq.(2.10))}$$

### *A.2. Equations for the system density matrix (eqs.(2.11), (2.12) and (2.13)):*

Next we derive a "transport equation" for the reduced density matrix defined as

$$\rho_{IJ}(t,t') \equiv \sum_R \tilde{\psi}_{IR}(t)\,\tilde{\psi}_{JR}(t')^*,$$

$$\text{that is, } [\rho(t,t')] \equiv \sum_R \{\psi(t)\}_R\,\{\psi(t')\}_R^+ \tag{A.13}$$

For this we look at a typical subspace associated with the contact being in state 'R' and consider the inflow and outflow to and from this subspace:

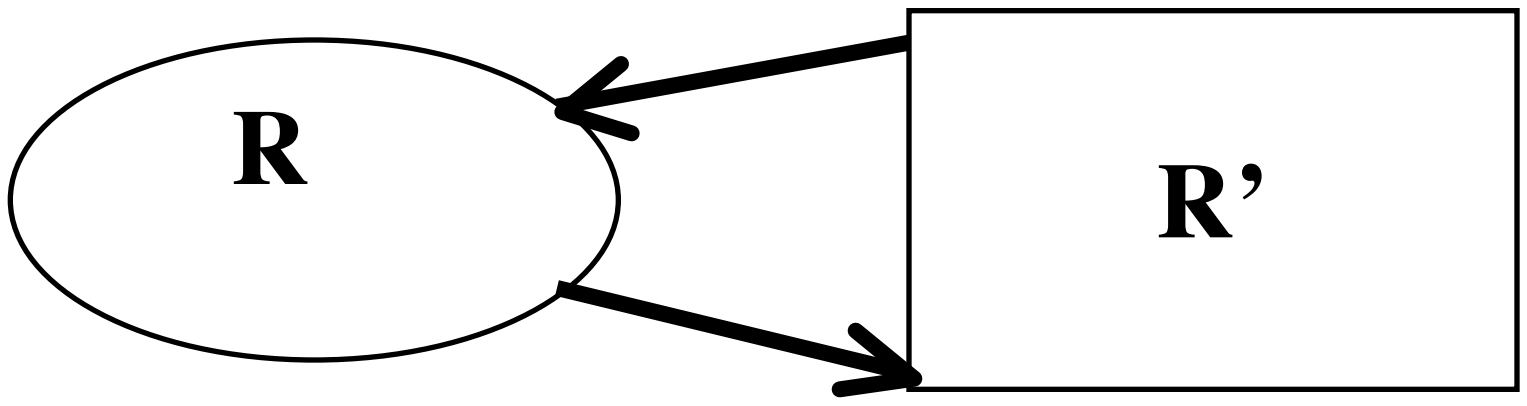

This can be described by an equation of the form

$$\left(i\hbar\frac{d}{dt}[I] - [H]\right)\{\psi(t)\}_R - \int_{-\infty}^{+\infty} dt_1\,[\Sigma(t,t_1)]\,\{\psi(t_1)\}_R = \{s(t)\}_R \tag{A.14}$$

$$\text{where } \{s(t)\}_R = \sum_{R'} [\tau(t)]_{R,R'}\{\psi(t)\}_{R'} \tag{A.15}$$

The source term {s} on the right of Eq.(A.14) represents the inscattering from other subspaces R' as evident from eq.(A.15), while the outflow occurs through the imaginary part of the self-energy. Making use of the Green's function defined in eq.(A.4), we can write the solution to eq.(A.14) in integral form

$$\{\psi(t)\}_R = \int_{-\infty}^{+\infty} dt_1\,[G(t,t_1)]\{s(t_1)\}_R \tag{A.16}$$

from which we write the density matrix in the form

$$\rho(t,t') = \sum_R \{\psi(t)\}_R \left\{\psi(t')^+\right\}_R$$

$$= \int_{-\infty}^{+\infty} dt_2 \int_{-\infty}^{+\infty} dt_1\, [G(t,t_2)]\{S(t_2,t_1)\}[G^+(t_1,t')] \qquad \text{(A.17, same as eq.(2.11))}$$

where

$$S(t,t') = \sum_R \{s(t)\}_R \left\{s(t')^+\right\}_R$$

$$= \sum_{R,R',R''} [\tau(t)]_{R,R'} \{\psi(t)\}_{R'} \left\{\psi^+(t')\right\}_{R''} [\tau(t')]_{R'',R} \qquad \text{(A.18)}$$

Assuming $\quad \psi_{KR'}(t)\psi_{LR''}(t')^* = P_{R'}\,\delta_{R'R''}\,\rho_{KL}(t,t') \qquad$ (A.19)

we obtain an equation relating the "source correlation function" S(t, $t'$) to the density matrix $\rho$(t, $t'$):

$$S_{MN}(t,t') = \sum_{K,L} f_{MNKL}(t,t')\,\rho_{KL}(t,t') \qquad \text{(A.20, same as eq.(2.12))}$$

where

$$f_{MNKL}(t,t') = \sum_{R,R'} \left[\tau(t)\right]_{MR,KR'} P_{R'} \left[\tau(t')\right]_{LR',NR}$$

$$= Trace_R\left[\rho_R\right]\left[\tau(t')\right]_{LN}\left[\tau(t)\right]_{MK}$$

$$\equiv \left\langle \left[\tau(t')\right]_{LN}\left[\tau(t)\right]_{MK} \right\rangle_R \qquad \text{(A.21, same as eq.(2.13))}$$

Note that there are two functions appearing in this formulation that describe the properties of each contact or reservoir: $F_{MNKL}(t,t')$ in eq.(A.11 or 2.10) and $f_{MNKL}(t,t_1)$ in eq.(A.21 or 2.13). These expressions are obtained using the "self-consistent Born approximation" and could in principle be evaluated using other approximations as well. However, we will show below that they have to satisfy the relationship stated in eq.(2.25) in order to ensure that inflow and outflow are balanced and probability is conserved.

### *A.3. Equations for the system observables (eqs.(2.14), (2.15) and (2.16)):*

Corresponding to any observable quantity 'Q' there is an operator $\Theta_Q$ in terms of which we can define a correlation function

$$Q(t,t') = \sum_R \{\psi(t')\}_R^+ \Theta_Q \{\psi(t)\}_R$$

$$= \text{Trace } [\rho(t,t')][\Theta_Q] \qquad \text{(A.22)}$$

such that the expectation value of 'Q' is given by $Q(t,t)$:

$$\langle Q(t) \rangle = \text{Q(t, t )} = Trace\,[\rho(t,t)]\,[\Theta_Q] \qquad \text{(A.23, same as eq.(2.14))}$$

We now associate a "current" $I_Q(t,t')$ with any quantity 'Q'

$$I_Q(t,t') = \left(\frac{d}{dt} + \frac{d}{dt'}\right) Q(t,t') \qquad \text{(A.24)}$$

such that $I_Q(t,t) = d\langle Q(t) \rangle / dt$ represents the rate of change of that quantity. For example, if 'Q' is the charge (= e * number of electrons) , then $I_Q$ is the usual current. We can obtain an expression for $I_Q$ that does not involve time derivatives by substituting eq.(A.22) into eq.(A.24)

$$I_Q(t,t') = \sum_R \{\psi(t')\}_R^+ \Theta_Q \left\{\frac{d\psi(t)}{dt}\right\}_R + \left\{\frac{d\psi(t')}{dt'}\right\}_R^+ \Theta_Q \{\psi(t)\}_R$$

and replacing the terms $d\psi(t)/dt$ and $d\psi(t')^+/dt'$ from eq.(A.14):

$$I_Q(t,t') = \frac{1}{i\hbar} \sum_R \{\psi(t')\}_R^+ \Theta_Q H\{\psi(t)\}_R - \{\psi(t')\}_R^+ H\Theta_Q\{\psi(t)\}_R$$

$$+ \int dt_1 \{\psi(t')\}_R^+ \Theta_Q \Sigma(t,t_1)\{\psi(t_1)\}_R - \int dt' \{\psi(t_1)\}_R^+ \Sigma^+(t_1,t')]\Theta_Q \{\psi(t)\}_R$$

$$+ \{\psi(t')\}_R^+ \Theta_Q \{s(t)\}_R - \{s(t')\}_R^+ \Theta_Q \{\psi(t)\}_R$$

Using eq.(A.16) to replace $\{\psi(t)\}$ in terms of {s(t)} in the last term and noting that the answer is a single number and hence can be written as a trace (this allows us to rearrange the matrices whose product is being traced) we obtain

$$\begin{aligned} I_Q(t,t') \;=\; & \frac{1}{i\hbar} Trace \sum_R \{\psi(t)\}_R \{\psi(t')\}_R^+ [\Theta_Q H - H\Theta_Q] \\ & + \int dt_1 \quad \Sigma(t,t_1)\{\psi(t_1)\}_R \{\psi(t')\}_R^+ \Theta_Q \;-\; \{\psi(t)\}_R \{\psi(t_1)\}_R^+ \Sigma^+(t_1,t')\Theta_Q \\ & + \int dt_1 \quad \{s(t)\}_R \{s(t_1)\}_R^+ G^+(t_1,t')\Theta_Q \;-\; G(t,t_1)\{s(t_1)\}_R \{s(t')\}_R^+ \Theta_Q \end{aligned}$$

Making use of the definitions of the density matrix (eq.(A.13)) and the source correlation function (eq.(A.18)), we obtain

$$I_Q(t,t') \;=\; \frac{1}{i\hbar} Trace \begin{pmatrix} [\rho(t,t')][\Theta_Q H - H\Theta_Q] \\ +\int dt_1 \, [\Sigma(t,t_1)][\rho(t_1,t')]\Theta_Q - [\rho(t,t_1)][\Sigma^+(t_1,t)]\Theta_Q \\ +\int dt_1 \; [S(t,t_1)][G^+(t_1,t')]\Theta_Q - [G(t,t_1)][S(t_1,t')]\Theta_Q \end{pmatrix} \qquad \text{(A.25)}$$

Setting $t' = t$, we obtain the relation stated earlier in eq.(2.15).

***Correlation functions:*** Finally we obtain an expression for the expectation values of quantities like $\langle c_i^+(t) c_j(t') \rangle$ where the operators $c_i^+(t)$ and $c_j(t')$ are expressed in the Heisenberg picture. In the Schrodinger picture (which is what we are using for the system and the interaction) we can write

$$\begin{aligned} \langle P(t)Q(t') \rangle \;&=\; \sum_R \{\psi(t)\}_R^+ \Theta_P U(t,t') \, \Theta_Q \{\psi(t')\}_R \\ &=\; Trace\, [\rho(t',t)\Theta_P U(t,t') \, \Theta_Q] \end{aligned} \qquad \text{(A.26, same as 2.16))}$$

where the matrix elements of the time evolution operator $U_{IJ}$(t, $t'$) represents the amplitude for the system to be in state $|I\rangle$ at time 't' if it was in state $|J\rangle$ at time $t'$. This is given by iG (t, $t'$), where G is the retarded Green's function in eq.(A.17) if t $> t'$, and by its Hermitian time-reversed conjugate -i$G^+$(t, $t'$) if t $< t'$.

$$U_{IJ}(t,t') = \ i\,[G(t,t') - G^{+}(t,t')]_{IJ} \quad = \quad i\,[G_{IJ}(t,t') - G_{JI}{}^{*}(t',t)] \qquad \text{(A.27)}$$

***A.4. Condition for conservation of probability (eqs.(2.25a, b)):*** We will now show that the total probability is conserved in our formalism provided

$$F_{MNKL}(t,t') \quad = \quad f_{LKNM}(t',t) \qquad \text{(A.28, same as 2.25a))}$$

which is ensured in the approximation used in this paper (see Eqs.(2.10) and (2.13)). To show eq.(A.28) we note that the total probability is given by eq.(2.14) with $\Theta_Q$ = I, so that the rate at which it changes is obtained from eq.(2.15):

$$\frac{i}{\hbar}\int_{-\infty}^{+\infty} dt_1\ Trace\begin{pmatrix}[G(t,t_1)]\,[S(t_1,t)] - [S(t,t_1)][G^{+}(t_1,t)] \\ +[\rho(t,t_1)][\Sigma^{+}(t_1,t)] - [\Sigma(t,t_1)][\rho(t_1,t)]\end{pmatrix}$$

$$= \ \frac{1}{\hbar}\,\mathrm{Re}\,al\int_{-\infty}^{+\infty} dt_1\ Trace\left([G(t,t_1)]\,[S(t_1,t)] - [\Sigma(t,t_1)][\rho(t_1,t)]\right)$$

which we can rewrite using Eqs.(2.9) and (2.12) as

$$\frac{1}{\hbar}\,\mathrm{Re}\,al\int_{-\infty}^{+\infty} dt_1 \sum_{K,L,M,N}\begin{pmatrix}G_{KL}(t,t_1)\,f_{LKNM}(t_1,t)\,\rho_{NM}(t_1,t) \\ -\,F_{MNKL}(t,t_1)\,G_{KL}(t,t_1)\,\rho_{NM}(t_1,t)\end{pmatrix}$$

which is equal to zero if Eq.(2.25a) is true.